\documentclass[a4paper,11pt]{amsart}
\usepackage{amssymb,amsmath,amsthm,mathrsfs}
\usepackage[colorlinks,linkcolor=blue,anchorcolor=blue,citecolor=blue,urlcolor=blue]{hyperref}
=15pt
\newtheorem{thm}{Theorem}
\newtheorem{lem}{Lemma}
\newtheorem{defn}{Definition}
\newtheorem{exa}{Example}
\newtheorem{rem}{Remark}

\numberwithin{equation}{section}

\begin{document}

\title[]{Several classes of bent, near-bent and 2-plateaued functions over
finite fields of odd characteristic}%
\author{Guangkui Xu, Xiwang Cao}%
\address{G. Xu is with the Department of Mathematics, Nanjing University of Aeronautics and Astronautics, Nanjing 210016, China,  and also with the Department of Mathematics, Huainan Normal University, Huainan 232038, China (e-mail: {\tt xuguangkuiy@163.com}).}
\address{Xiwang Cao is with the School of Mathematical Sciences, Nanjing University of
Aeronautics and Astronautics, Nanjing 210016, China, and also with the State Key Laboratory of Information security, Institute of Information Engineering, Chinese Academy of Sciences, Beijing 100093, China (email: {\tt
xwcao@nuaa.edu.cn}).}


\begin{abstract}
Inspired by a recent work of Mesnager, we present several new infinite families of quadratic ternary  bent,  near-bent and   2-plateaued functions from some  known quadratic ternary bent functions. Meanwhile, the distribution of the Walsh spectrum of two class of 2-plateaued functions obtained in this paper is  completely
determined. Additionally, we  construct the first class of $p$-ary bent functions of algebraic degree $p$ over the fields of an arbitrary odd characteristic. The proposed class contains  non-quadratic $p$-ary bent functions that are affinely inequivalent
to  known monomial and binomial ones.
\end{abstract}
\maketitle

\section{Introduction}

Boolean functions with few Walsh transform values have useful applications in cryptography and communications.
Such functions provide protection against linear cryptanalysis in cryptography \cite{matsui1994linear} and correspond to sequences that have low cross-correlation
with the $m$-sequence in communications \cite{helleseth1999correlation}. Boolean bent functions which  have the highest possible nonlinearity with even number of variables were first introduced by Rothaus \cite{rothaus1976bent} in 1976. Bent functions have been  widely studied and have received much attention due to their significantly important role in cryptography, coding theory, communication, and sequence design. As a generalization of Rothau's  notion of  a bent function, Kumar, Scholtz and Welch  extended it to $p$-ary bent functions from $\mathbb{Z}_p^n$ to $\mathbb{Z}_{p}$ in \cite{kumar1985generalized}, where $p$ is an integer. In the case of   $p$-ary bent functions things are naturally much more complicated compare to Boolean bent functions.
A number of recent papers are devoted to the description of new classes of  bent functions.  However, there are a few known families of bent functions. In general, there are two ways to construct bent functions: primary constructions and secondary constructions. For some primary and secondary constructions of bent functions on monomials, binomials and quadratic functions, the reader is referred to  \cite{canteaut2008new,carlet1994two, carlet2011dillon's, dillon1974elementary, helleseth2010new, li2013several, li2008constructions, mesnager2011new}. Surveys of known results on  bent functions  can be
seen in \cite{carlet2010boolean,dobbertin2005survey,mesnager} and the references therein. Plateaued functions were introduced  by Zheng and Zhang
as good candidates for designing cryptographic functions since they possess
desirable various cryptographic characteristics \cite{zhengp}. For
more results on the treatment of plateaued functions, we refer to \cite{Prouff,cesmelioglu, cesmelioglu1, charpin2005bent,chee, khoo2006new, Mesnagerbook,zheng1,zheng1999}


   For a prime $p$ and a positive integer $n$, let $ \mathbb{F}_{p^n}$ be the finite field with $p^n$ elements and $\mathbb{F}_{p^n}^{\ast}=\mathbb{F}_{p^n}\setminus\{0\}$. For any $k$ dividing $n$, we denote the trace function from $\mathbb{F}_{p^{n}}$ to
$\mathbb{F}_{p^{k}}$ as follows:
$${\rm Tr}_{k}^{n}(x)=x+x^{p^{k}}+x^{p^{2k}}+\cdots+x^{p^{n-k}}.$$
For $k=1$, ${\rm Tr}_{1}^{n}(x)=\sum\limits_{i=0}^{n-1}x^{p^{i}}$ is called the absolute trace function.  Recently, Mesnager \cite{mesnager2014several}   proved a strong version of \cite[Theorem 3]{carlet2006bent}, and provided several primary and secondary
constructions of bent functions. Via some known binary monomial bent functions and their corresponding  dual functions, she obtained two new infinite families of bent functions with the forms
\begin{equation}\label{e1}
f(x)={\rm Tr}_{1}^{k}(\lambda x^{2^k+1})+{\rm Tr}_{1}^{n}(ux){\rm Tr}_{1}^{n}(vx)
\end{equation}
and
\begin{equation}\label{ej2}
f(x)={\rm Tr}_{1}^{k}( x^{2^k+1})+{\rm Tr}_{1}^{n}\bigg( \sum_{i=1}^{2^{r-1}-1}x^{(2^k-1)\frac{i}{2^r}+1}\bigg)+{\rm Tr}_{1}^{n}(ux){\rm Tr}_{1}^{n}(vx)
\end{equation}
over $\mathbb{F}_{2^{n}}$, where $n=2k$, $\lambda\in \mathbb{F}_{2^{k}}^*$ and $u,v \in \mathbb{F}_{2^{n}}^*$. It is well known that a function given by adding a linear function to one  bent function is also bent. Inspired by the work of \cite{mesnager2014several},  a very natural question is considered:
whether we can obtain  new $p$-ary bent functions by adding the product of two linear functions to some known $p$-ary bent functions? In a very recent paper \cite{xuguangkui}, the authors proved that some such $p$-ary functions are also bent under certain conditions. In this paper, we will continue the work of \cite{mesnager2014several,xuguangkui} and construct   more ternary bent  and 2-plateaued functions of the form
\begin{equation}\label{ej4}
f(x)={\rm Tr}_{1}^{k}(\lambda x^{3^k+1})+{\rm Tr}_{1}^{n}(ux){\rm Tr}_{1}^{n}(vx)
\end{equation}
over $\mathbb{F}_{3^{n}}$, where $n=2k$, $\lambda\in \mathbb{F}_{3^{k}}^*$ and $u,v \in \mathbb{F}_{3^{n}}^*$. For an odd prime $p$, some $p$-ary bent functions of the form
\begin{equation}\label{ej3}
f(x)={\rm Tr}_{1}^{n}(\lambda x^{d})-{\rm Tr}_{1}^{n}(ux){\rm Tr}_{1}^{n}(x)^{p-1}
\end{equation}
over $\mathbb{F}_{p^{n}}$ are obtained, where $d=p^k+1$ or $2$.
The main result of this paper is obtained from the study of the Walsh transform, which is different from the one used in \cite{mesnager2014several}.

The rest of the paper is organized as follows. In Section 2, we give some notations and related
results. In Section 3, using some known ternary  bent functions, we derive more ternary bent, near-bent and 2-plateaued functions of the form (\ref{ej4}).  In Section 4, two new classes of non-quadratic $p$-ary bent functions of algebraic degree $p$ with the form (\ref{ej3}) are obtained.

\section{Preliminaries }

A polynomial $f(x)\in \mathbb{F}_{p^{n}} [x]$ is called a {\it permutation polynomial}  of $\mathbb{F}_{p^{n}}$ if it induces a bijective map  from $\mathbb{F}_{p^{n}}$ to $\mathbb{F}_{p^{n}}$.  For a permutation  polynomial $f(x)\in \mathbb{F}_{p^{n}} [x]$ there exists  (a unique) $ f^{-1}(x)\in \mathbb{F}_{p^{n}} [x]$ such that $f(f^{-1}(x))\equiv f^{-1}(f(x))\equiv x\  ({\rm mod}\  x^{p^{n}}-x)$. We call $f^{-1}(x)$ the {\it compositional inverse} of $f(x)$.
 A polynomial of the form
$$L(x)=\sum\limits_{i=0}^{m}a_ix^{p^i} \in \mathbb{F}_{p^{n}}[x]$$
is called a {\it linearized polynomial}. It is well known that the  compositional inverse of a linearized permutation polynomial is  also a linearized polynomial. Recently, the compositional inverse of of some  linearized permutation polynomials have been discovered. For detailed information, the reader is referred to \cite{Coulter, Tuxanidy, wu, wu2013linearized}.

For some $l\mid n$,  denote by $N_{l}^{n}:\mathbb{F}_{p^n}\rightarrow \mathbb{F}_{p^l}$ the norm function $N_{l}^{n}(x)=x^{\frac{p^n-1}{p^l-1}}$.
{\lem [Theorem 2.1 \cite{wu}] \label{pppl1} Let $a\in \mathbb{F}_{p^{n}}^*$ and $d=\gcd(n,r)$. Then the linearized
binomial $L_{a,r}(x)=x^{p^r}+ax$ is a permutation polynomial over  $\mathbb{F}_{p^{n}}$ if and only if $(-1)^{\frac{n}{d}}N_{d}^{n}(a)\neq1$. Moreover,
$$L_{a,r}^{-1}(x)=\frac{N_{d}^{n}(a)}{N_{d}^{n}(a)+(-1)^{\frac{n}{d}}}
\sum_{i=0}^{\frac{n}{d}-1}(-1)^ia^{-\frac{p^{(i+1)r}-1}{p^r-1}}x^{p^{ir}}.$$
In particular, if $n$ is even, then $L_{a,\frac{n}{2}}(x)=x^{p^{\frac{n}{2}}}+ax$ is a linearized permutation
polynomial over   $\mathbb{F}_{p^{n}}$ if and only if $a^{p^{\frac{n}{2}}+1}\neq1$, and
$$L_{a,\frac{n}{2}}^{-1}(x)=\frac{1}{a^{p^{\frac{n}{2}}+1}-1}(a^{p^{\frac{n}{2}}}x-x^{\frac{n}{2}}).$$}

Let $f:\mathbb{F}_{p^{n}}
\rightarrow \mathbb{F}_{p}$ be a $p$-ary function with $n$ variables.
For any integer $d\in \{0,1,\cdots, p^n-1\}$, let $\sum_{i=0}^{n-1}d_ip^i$ be its $p$-ary expansion with $0\leq d_i\leq p-1$, then  the $p$-{\it weight} of $d$  denoted by
$w_p( d )$ equals $\sum_{i=0}^{n-1}d_i$.   Recall that each
function $f: \mathbb{F}_{p^{n}}\rightarrow\mathbb{F}_{p}$ can be represented by a univariate polynomial over
$\mathbb{F}_{p^{n}}$. The algebraic degree of $f (x)$ is equal to the maximum $p$-weight
of an exponent $j$  of the term $a_jx^j$ in $f (x)$ with $a_j\neq0$.

  The Walsh transform of $f$ is defined as follows
\begin{eqnarray*}\label{e4}
 \widehat{\chi}_f(a)=\sum_{x\in\mathbb{F}_{p^{n}}}\omega^{f(x)-{\rm Tr}^{n}_{1}(a
x)}, a\in\mathbb{F}_{p^{n}},
\end{eqnarray*}
 where $\omega=e^{2\pi \sqrt{-1}/p}$ is a primitive $p$-th root of unity.
The values $\widehat{\chi}_f(a), a \in\mathbb{F}_{p^{n}}$  are called the {\it Walsh coefficients} of $f$. The {\it Walsh spectrum} of a Boolean function $f$ is the multiset $\{\widehat{\chi}_f(a), a \in\mathbb{F}_{p^{n}}\}$.
 {\defn A   $p$-ary function $f$ is called  {\it  bent}  if $|
 \widehat{\chi}_f(a)|=p^{n/2}$ for all
 $a\in\mathbb{F}_{p^{n}}$.}

A $p$-ary bent function $f(x)$ is called {\it regular} if $
 \widehat{\chi}_f(a)=p^{n/2}\omega^{f^*(a)}$ for any
 $a\in\mathbb{F}_{p^{n}}$, where the function $f^*(x)$ is called the dual of $f(x)$.
 A bent function $f (x)$ is called {\it weakly
regular} if there is a complex $\mu$ with unit magnitude such that $
 \widehat{\chi}_f(a)=p^{n/2}\mu \omega^{f^*(a)}$.   The dual of a
(weakly) regular bent function is again a (weakly) regular bent
function \cite{helleseth2006monomial}.

{\defn \cite{cesmelioglu} For an integer $0\leq s \leq n$,  if $|\widehat{\chi}_f(a)|=\{0,p^{\frac{n+s}{2}}\}$ for all
 $a\in\mathbb{F}_{p^{n}}$, then we call $f$ {\it $s$-plateaued} (for $s=1$ the term {\it near-bent} is common).}

For $s\in \{0,1,2\}$, $s$-plateaued functions have been actively studied and have
attractive much attention due to their cryptographic, algebraic, and combinatorial
properties.


Two  functions $f,g:\mathbb{F}_{p^{n}}\rightarrow\mathbb{F}_{p}$ are called {\it affinely equivalent} {\cite{jia2012class,zheng} if
$f (x) = ag(l(x) + b) + c$ for some linearized permutation $l(x) \in \mathbb{F}_{p^{n}}[x]$, $a, c \in\mathbb{F}_{p}$ and  $b \in \mathbb{F}_{p^{n}}$. Note that algebraic degree, bentness of a $p$-ary function are affine
invariants.

In this paper, we mainly focus on the secondary constructions  via some known $p$-ary bent functions. Thus, we need to recall the following known results.

{\lem [$p$-ary Kasami, Corollary 4 \cite{liu1992nonbinary, helleseth2006monomial}]\label{pl1} Let $n=2k$ and
$\lambda \in \mathbb{F}_{p^{k}}^*$. For any odd prime $p$, the $p$-ary monomial $f(x)={\rm Tr}^{k}_{1}(\lambda x^{p^k+1})$
is a weakly regular bent function.   Moreover, for $a\in \mathbb{F}_{p^{n}}$ the corresponding Walsh transform coefficient of $f(x)$ is equal to $$\widehat{\chi}_f(a)=-p^k\omega^{-{\rm Tr}^{k}_{1}\left(\lambda^{-1}a^{p^k+1}\right)}.$$ }
{\lem [Sidelnikov, Corollary 3 \cite{helleseth2006monomial}]\label{l3} For any nonzero
$\lambda \in \mathbb{F}_{p^{n}}$ and  odd prime $p$, the $p$-ary monomial $f(x)={\rm Tr}^{n}_{1}(\lambda x^{2})$
is a (weakly) regular bent function. Moreover, for $a\in \mathbb{F}_{p^{n}}$ the corresponding Walsh transform coefficient of $f(x)$ is equal to
\begin{equation}\label{e5}
\widehat{\chi}_f(a)=\left\{
  \begin{array}{ll}
    \eta(\lambda)(-1)^{n-1}p^{\frac{n}{2}}\omega^{-{\rm Tr}^{n}_{1}(\frac{a^2}{4\lambda})}, & \hbox{if $p\equiv 1 \ ({\rm mod} 4)$;} \\
    \eta(\lambda)(-1)^{n-1}i^np^{\frac{n}{2}}\omega^{-{\rm Tr}^{n}_{1}(\frac{a^2}{4\lambda})}, & \hbox{if $p\equiv 3 \ ({\rm mod} 4)$,}
  \end{array}
\right.
\end{equation}
where $i=\sqrt{-1}$ and $\eta$ is the quadratic character of $\mathbb{F}_{p^{n}}$.}


\section{Quadratic ternary bent, near-bent and 2-plateaued functions from some known  ternary bent functions }

 In this section, we will  construct several  classes of quadratic ternary bent, near-bent and 2-plateaued functions. Before doing this, we need the following lemma.

{\lem [ Lemma 4 \cite{xuguangkui}]\label{l5} Let $n$ be a positive integer and $u,v \in \mathbb{F}_{3^{n}}^*$. Let $g(x)$ be a ternary function defined on $\mathbb{F}_{3^{n}}$. Define  the ternary function $f(x)$ by $$f(x)=g(x)+{\rm Tr}_{1}^{n}(ux){\rm Tr}_{1}^{n}(vx).$$ Then for any $a\in \mathbb{F}_{3^{n}}$, the corresponding Walsh transform coefficient of $f(x)$ is equal to
\begin{align}
 \widehat{\chi}_f(a)=&\frac{1}{3}[\widehat{\chi}_g(a)+\widehat{\chi}_g(a+u)+ \widehat{\chi}_g(a-u)
 \notag\\
&+\widehat{\chi}_g(a-v)+\omega\widehat{\chi}_g(a-v+u)+ \omega^2\widehat{\chi}_g(a-v-u)
\notag\\&+\widehat{\chi}_g(a+v)+\omega^2\widehat{\chi}_g(a+v+u)+ \omega\widehat{\chi}_g(a+v-u)],\label{e6}
\end{align}
where $\omega$ is a primitive $3$-rd root of unity. }

It was shown in \cite{cesmelioglu1} that quadratic monomial near-bent functions ${\rm Tr}^{n}_{1}(\lambda x^{p^r+1}),$ $\lambda \in \mathbb{F}_{p^n}$  in odd characteristic $p$ do not exist. Now, we can derive the following construction  of ternary near-bent functions via ternary monomial bent function ${\rm Tr}^{n}_{1}(\lambda x^{3^r+1})$.
We define a subset of $\mathbb{F}_3^{3}$ as
$$A=\{(0,1,1), (0,2,2),(1,1,1),(1,2,2),(2,0,1),( 2,0,2),(2,1,0),(2,2,0)\}.$$
Now, using Lemmas \ref{pl1} and \ref{l5}, we can obtain the next theorem.

{\thm \label{t1}Let $k>1$ and $n=2k$. Let  $\lambda\in \mathbb{F}_{3^{k}}^*$ and  $u, v \in \mathbb{F}_{3^{n}}^* $. Let $f$ be the ternary  function defined as
\begin{eqnarray}\label{ej12}
f(x)={\rm Tr}^{k}_{1}(\lambda x^{3^k+1})+{\rm Tr}_{1}^{n}(ux){\rm Tr}_{1}^{n}(vx).
\end{eqnarray}
\begin{description}
  \item[1)]  If $({\rm Tr}^{n}_{1}(\lambda^{-1}u^{3^k}v),{\rm Tr}^{k}_{1}(\lambda^{-1}u^{3^k+1}),{\rm Tr}^{k}_{1}(\lambda^{-1}v^{3^k+1}))\in \mathbb{F}_3^{3}\setminus (A\cup\{(2,0,0)\})$,
  then $f$ is bent.
  \item[2)] If $({\rm Tr}^{n}_{1}(\lambda^{-1}u^{3^k}v),{\rm Tr}^{k}_{1}(\lambda^{-1}u^{3^k+1}),{\rm Tr}^{k}_{1}(\lambda^{-1}v^{3^k+1}))\in A$,
  then $f$ is near-bent.

  \item[3)] If $({\rm Tr}^{n}_{1}(\lambda^{-1}u^{3^k}v),{\rm Tr}^{k}_{1}(\lambda^{-1}u^{3^k+1}),{\rm Tr}^{k}_{1}(\lambda^{-1}v^{3^k+1}))=(2,0,0)$, then
   $f$ is a ternary 2-plateaued function. Moreover, when $a$ runs through all elements in $\mathbb{F}_{3^{n}}$, the distribution of the
Walsh spectrum of    ternary 2-plateaued function $f$  is given by
\begin{eqnarray*}
\widehat{\chi}_f(a)=\left\{
  \begin{array}{ll}
    0, & \hbox{occurs $3^{n}-3^{n-2}$ times,} \\
    -3^{k+1}, & \hbox{occurs $3^{n-3}-2\cdot3^{k-2}$ times,} \\
    -3^{k+1}\omega, & \hbox{occurs $3^{n-3}+3^{k-2}$ times,} \\
    -3^{k+1}\omega^2, & \hbox{occurs $3^{n-3}+3^{k-2}$ times}.
  \end{array}
\right.
\end{eqnarray*}

\end{description}}

\begin{proof}
Let   $g(x)={\rm Tr}^{k}_{1}(\lambda x^{3^k+1})$.
For $a\in \mathbb{F}_{3^{n}}$, it follows from  Lemma \ref{l5} that
\begin{eqnarray*}
 \widehat{\chi}_f(a)=\triangle_1+ \triangle_3+\triangle_3,
\end{eqnarray*}
where
\begin{align*}
\triangle_1=\frac{1}{3}[\widehat{\chi}_g(a)+\widehat{\chi}_g(a+u)+ \widehat{\chi}_g(a-u)],
\end{align*}
\begin{align*}
\triangle_2=&\frac{1}{3}[\widehat{\chi}_g(a-v)+\omega\widehat{\chi}_g(a-v+u)+\omega^2 \widehat{\chi}_g(a-v-u)]
\end{align*}
and
\begin{align*}
\triangle_3=&\frac{1}{3}[\widehat{\chi}_g(a+v)+\omega^2\widehat{\chi}_g(a+v+u)+\omega \widehat{\chi}_g(a+v-u)].
\end{align*}
From Lemma \ref{pl1}, we have   $ \widehat{\chi}_g(a)=-3^k\omega^{-{\rm Tr}^{k}_{1}(\lambda^{-1}a^{3^k+1})}.$
  Then
\begin{align*}
\triangle_1=&\frac{1}{3}[\widehat{\chi}_g(a)+\widehat{\chi}_g(a+u)+ \widehat{\chi}_g(a-u)]
\\=&\frac{1}{3}[-3^k\omega^{-{\rm Tr}^{k}_{1}(\lambda^{-1}a^{3^k+1})}
 -3^k\omega^{-{\rm Tr}^{k}_{1}(\lambda^{-1}a^{3^k+1})
-{\rm Tr}^{k}_{1}(\lambda^{-1}(a^{3^k}u+au^{3^k}))-{\rm Tr}^{k}_{1}(\lambda^{-1}u^{3^k+1})}
\\&-3^k\omega^{-{\rm Tr}^{k}_{1}(\lambda^{-1}a^{3^k+1})
+{\rm Tr}^{k}_{1}(\lambda^{-1}(a^{3^k}u+au^{3^k}))-{\rm Tr}^{k}_{1}(\lambda^{-1}u^{3^k+1})}
]
\\=&-\frac{1}{3}3^k\omega^{-{\rm Tr}^{k}_{1}(\lambda^{-1}a^{3^k+1})}[1+\omega^{
-{\rm Tr}^{n}_{1}(\lambda^{-1}a^{3^k}u)-{\rm Tr}^{k}_{1}(\lambda^{-1}u^{3^k+1})}
\\&+\omega^{
{\rm Tr}^{n}_{1}(\lambda^{-1}a^{3^k}u)-{\rm Tr}^{k}_{1}(\lambda^{-1}u^{3^k+1})}]
\end{align*}
where the last identity holds due to the transitivity property of ${\rm Tr}^{n}_{1}(x)$, that is, ${\rm Tr}^{n}_{1}(x)={\rm Tr}^{k}_{1}({\rm Tr}^{n}_{k}(x))$.

Similarly, we have
\begin{align*}
\triangle_2=&\frac{1}{3}[(\widehat{\chi}_g(a-v)+\omega\widehat{\chi}_g(a-v+u)+\omega^2 \widehat{\chi}_g(a-v-u)]
\\=&-\frac{1}{3}3^k\omega^{-{\rm Tr}^{k}_{1}(\lambda^{-1}a^{3^k+1})+{\rm Tr}^{n}_{1}(\lambda^{-1}a^{3^k}v)-{\rm Tr}^{k}_{1}(\lambda^{-1}v^{3^k+1})}
\\&\times[1+
\omega^{1-{\rm Tr}^{n}_{1}(\lambda^{-1}a^{3^k}u)+{\rm Tr}^{n}_{1}(\lambda^{-1}u^{3^k}v)-{\rm Tr}^{k}_{1}(\lambda^{-1}u^{3^k+1})}
\\&+\omega^{
2+{\rm Tr}^{n}_{1}(\lambda^{-1}a^{3^k}u)-{\rm Tr}^{n}_{1}(\lambda^{-1}u^{3^k}v)-{\rm Tr}^{k}_{1}(\lambda^{-1}u^{3^k+1})}]
\end{align*}
and
\begin{align*}
\triangle_3=&\frac{1}{3}[\widehat{\chi}_g(a+v)+\omega^2\widehat{\chi}_g(a+v+u)+\omega \widehat{\chi}_g(a+v-u)]
\\=&-\frac{1}{3}3^k\omega^{-{\rm Tr}^{k}_{1}(\lambda^{-1}a^{3^k+1})-{\rm Tr}^{n}_{1}(\lambda^{-1}a^{3^k}v)-{\rm Tr}^{k}_{1}(\lambda^{-1}v^{3^k+1})}
\\ &\times[1+
\omega^{2-{\rm Tr}^{n}_{1}(\lambda^{-1}a^{3^k}u)-{\rm Tr}^{n}_{1}(\lambda^{-1}u^{3^k}v)-{\rm Tr}^{k}_{1}(\lambda^{-1}u^{3^k+1})}
\\&+\omega^{
1+{\rm Tr}^{n}_{1}(\lambda^{-1}a^{3^k}u)+{\rm Tr}^{n}_{1}(\lambda^{-1}u^{3^k}v)-{\rm Tr}^{k}_{1}(\lambda^{-1}u^{3^k+1})}].
\end{align*}
Let $t_0={\rm Tr}^{n}_{1}(\lambda^{-1}u^{3^k}v)$, $t_1={\rm Tr}^{k}_{1}(\lambda^{-1}u^{3^k+1})$ and $t_2={\rm Tr}^{k}_{1}(\lambda^{-1}v^{3^k+1})$. Denote $c_1={\rm Tr}^{n}_{1}(\lambda^{-1}a^{3^k}u)$ and $c_2={\rm Tr}^{n}_{1}(\lambda^{-1}a^{3^k}v)$. Then the sums $\triangle_1$, $\triangle_2$ and $\triangle_3$ can be written as
\begin{eqnarray}\label{q-7}
\triangle_1=-\frac{1}{3}3^k\omega^{-{\rm Tr}^{k}_{1}(\lambda^{-1}a^{3^k+1})}(1+\omega^{
-c_1-t_1}+\omega^{
c_1-t_1}),
\end{eqnarray}
\begin{align}\label{q-8}
\triangle_2=&-\frac{1}{3}3^k\omega^{-{\rm Tr}^{k}_{1}(\lambda^{-1}a^{3^k+1})+c_2-t_2}(1+
\omega^{1-c_1+t_0-t_1}+\omega^{
2+c_1-t_0-t_1})
\end{align}
and
\begin{align}\label{q-9}
\triangle_3=&-\frac{1}{3}3^k\omega^{-{\rm Tr}^{k}_{1}(\lambda^{-1}a^{3^k+1})-c_2-t_2}(1+
\omega^{2-c_1-t_0-t_1}+\omega^{
1+c_1+t_0-t_1}).
\end{align}
Combining (\ref{q-7})-(\ref{q-9}), we get
\begin{align}\label{qqq-10}
\widehat{\chi}_f(a)=&\triangle_1+\triangle_2+\triangle_3
\notag\\=&-\frac{1}{3}3^k\omega^{-{\rm Tr}^{k}_{1}(\lambda^{-1}a^{3^k+1})}[1+\omega^{-c_1-t_1}+\omega^{c_1-t_1}
\notag\\&+\omega^{c_2-t_2}+\omega^{-c_2-t_2}+\omega^{1-c_1+c_2+t_0-t_1-t_2}
+\omega^{2+c_1+c_2-t_0-t_1-t_2}\notag\\&+\omega^{2-c_1-c_2-t_0-t_1-t_2}+\omega^{1+c_1-c_2+t_0-t_1-t_2}].
\end{align}

1) We only give the proof of the
case of $(t_0, t_1, t_2) = (1,2,1)\in\mathbb{F}_3^{3}\setminus (A\cup\{(2,0,0)\}) $ since the others can be proven
in a similar manner. Note that $(c_0,c_1)\in \mathbb{F}_3^2$ for a fixed $a\in\mathbb{F}_{3^n}$. Depending on  the value of the pair $(c_0,c_1)$ and from (\ref{qqq-10}), we have
\begin{align*}
\widehat{\chi}_f(a)&=\triangle_1+ \triangle_3+\triangle_3
\\&=\left\{
      \begin{array}{ll}
        3^k\omega^{-{\rm Tr}^{k}_{1}(\lambda^{-1}a^{3^k+1})}, & \hbox{if $(c_1,c_2)=(0,0)$,} \\
        3^k\omega^{-{\rm Tr}^{k}_{1}(\lambda^{-1}a^{3^k+1})+1}, & \hbox{if $(c_1,c_2)\in\{(1,0),(1,1),(2,0),(2,2)\},$} \\
        3^k\omega^{-{\rm Tr}^{k}_{1}(\lambda^{-1}a^{3^k+1})+2}, & \hbox{if $(c_1,c_2)\in\{(0,1),(0,2),(1,2),(2,1)\}$.}
      \end{array}
    \right.
\end{align*}
Hence, for all $a\in \mathbb{F}_{3^{n}}$, we can see that $|\widehat{\chi}_f(a)|=3^k$ if $(t_0, t_1, t_2) = (1,2,1)$.

2) Let $\omega=\frac{-1+\sqrt{3}i}{2}$ where $i=\sqrt{-1}$. It is easy to verify that $\omega^2-\omega=-\sqrt{3}i$. Similar as in 1), we only give  the proof of the
case of $(t_0, t_1, t_2) = (0,1,1)\in A$. For  $a\in \mathbb{F}_{3^{n}}$,  from (\ref{qqq-10}), we have
\begin{align*}
\widehat{\chi}_f(a)&=\triangle_1+ \triangle_3+\triangle_3
\\&=\left\{
      \begin{array}{ll}
        3^{k+1/2}i\omega^{-{\rm Tr}^{k}_{1}(\lambda^{-1}a^{3^k+1})}, & \hbox{if $(c_1,c_2)=(0,0)$,} \\
        3^{k+1/2}i\omega^{-{\rm Tr}^{k}_{1}(\lambda^{-1}a^{3^k+1})+1}, & \hbox{if $(c_1,c_2)\in\{(1,2),(2,1)\},$} \\
        0, & \hbox{otherwise.}
      \end{array}
    \right.
\end{align*}
Consequently,  for  $a\in \mathbb{F}_{3^{n}}$, $|\widehat{\chi}_f(a)|\in \{0,3^{k+1/2}\}$ if $(t_0, t_1, t_2) = (0,1,1)$.

3) If  $t_0=2$, $t_1=0$ and $t_2=0$, by (\ref{q-7})-(\ref{q-9}), we have
\begin{align}\label{pp13}
\widehat{\chi}_f(a)&=\triangle_1+ \triangle_3+\triangle_3
\notag\\&=-\frac{1}{3}3^k\omega^{-{\rm Tr}^{k}_{1}(\lambda^{-1}a^{3^k+1})}(1+\omega^{
-c_1}+\omega^{
c_1})(1+\omega^{
c_2}+\omega^{
-c_2})
\notag\\&=\left\{
      \begin{array}{ll}
        -3^{k+1}\omega^{-{\rm Tr}^{k}_{1}(\lambda^{-1}a^{3^k+1})}, & \hbox{if $(c_1,c_2)=(0,0)$,} \\
        0, & \hbox{otherwise.}
      \end{array}
    \right.
\end{align}
Therefore, $|\widehat{\chi}_f(a)|\in \{0,3^{k+1}\}$ for all $a\in \mathbb{F}_{3^{n}}$. Then $f$ is 2-plateaued.

By a complex and lengthy computation, we can obtain the value distribution of the Walsh
transform of 2-plateaued function $f$, and the details are presented in Appendix \ref{ap1}.
\end{proof}

{\exa Let $k=3$, $n=6$.  Let $\alpha$ be the generator of $\mathbb{F}_{3^6}^*$ with $\alpha^6 -\alpha^4 + \alpha^2 -\alpha - 2=0$.
\begin{description}
  \item[1)] Take $\lambda=\alpha^{84}$, $u=\alpha^4$ and $v=\alpha^6$. Then $f$ defined by (\ref{ej12}) is
$$f(x)={\rm Tr}^{3}_{1}(\alpha^{84} x^{28})+{\rm Tr}_{1}^{6}(\alpha^4x){\rm Tr}_{1}^{6}(\alpha^6x).$$
By a Magma program, we can see that  ${\rm Tr}_{1}^{6}(\lambda^{-1}u^{27}v)=1$, ${\rm Tr}_{1}^{3}(\lambda^{-1}u^{28})=0$, ${\rm Tr}_{1}^{3}(\lambda^{-1}v^{28})=0$  and $f$ is bent.
   \item[2)] Take $\lambda=\alpha^{84}$, $u=\alpha^7$ and $v=\alpha^{25}$. Then $f$ defined by (\ref{ej12}) is
$$f(x)={\rm Tr}^{3}_{1}(\alpha^{84} x^{28})+{\rm Tr}_{1}^{6}(\alpha^7x){\rm Tr}_{1}^{6}(\alpha^{25}x).$$
By a Magma program, we can see that  ${\rm Tr}_{1}^{6}(\lambda^{-1}u^{27}v)=2$, ${\rm Tr}_{1}^{3}(\lambda^{-1}u^{28})=2$, ${\rm Tr}_{1}^{3}(\lambda^{-1}v^{28})=0$ and $f$ is near-bent.

  \item[3)] Take $\lambda=\alpha^{84}$, $u=\alpha^4$ and $v=\alpha^{25}$. Then $f$ defined by (\ref{ej12}) is
$$f(x)={\rm Tr}^{3}_{1}(\alpha^{84} x^{28})+{\rm Tr}_{1}^{6}(\alpha^4x){\rm Tr}_{1}^{6}(\alpha^{25}x).$$
By a Magma program, we can see that  ${\rm Tr}_{1}^{6}(\lambda^{-1}u^{27}v)=2$, ${\rm Tr}_{1}^{3}(\lambda^{-1}u^{28})=0$, ${\rm Tr}_{1}^{3}(\lambda^{-1}v^{28})=0$ and $f$ is 2-plateaued. Moreover, the value distribution of the Walsh
transform of $f$ is
\begin{eqnarray*}
\widehat{\chi}_f(a)=\left\{
  \begin{array}{ll}
    0, & \hbox{occurs $648$ times,} \\
    -81, & \hbox{occurs $21$ times,} \\
    -81\omega, & \hbox{occurs $30$ times,} \\
    -81\omega^2, & \hbox{occurs $30$ times.}
  \end{array}
\right.
\end{eqnarray*}
\end{description}
Thus, our computer experiments are consistent with the results given in  Theorem \ref{t1}.}

We define a subset of $\mathbb{F}_3^{3}$ as
$$B=\{(0,1,1), (0,2,2),(1,0,1),(1,0,2),(1,1,0),( 1,2,0),(2,1,1),(2,2,2)\}.$$
Similarly, by Lemmas  \ref{l3} and \ref{l5}, we can prove the following theorem.
{\thm\label{t2}   Let $n$ be a positive integer with $n>3$   and  $\lambda,u, v \in \mathbb{F}_{3^{n}}^*$. The ternary  function $f$ defined as
\begin{eqnarray}\label{e12}
f(x)={\rm Tr}^{n}_{1}(\lambda x^{2})+{\rm Tr}_{1}^{n}(ux){\rm Tr}_{1}^{n}(vx).
\end{eqnarray}
\begin{description}
  \item[1)]  If $({\rm Tr}^{n}_{1}(\lambda^{-1}uv),{\rm Tr}^{n}_{1}(\lambda^{-1}u^{2}),{\rm Tr}^{n}_{1}(\lambda^{-1}v^{2}))\in \mathbb{F}_3^{3}\setminus (B\cup\{(1,0,0)\})$,
  then $f$ is bent.
  \item[2)] If $({\rm Tr}^{n}_{1}(\lambda^{-1}uv),{\rm Tr}^{n}_{1}(\lambda^{-1}u^{2}),{\rm Tr}^{n}_{1}(\lambda^{-1}v^{2}))\in B$,
  then $f$ is near-bent.

  \item[3)] If $({\rm Tr}^{n}_{1}(\lambda^{-1}uv),{\rm Tr}^{n}_{1}(\lambda^{-1}u^{2}),{\rm Tr}^{n}_{1}(\lambda^{-1}v^{2}))=(1,0,0)$, then
   $f$ is a ternary 2-plateaued function. Moreover, when $a$ runs through all elements in $\mathbb{F}_{3^{n}}$, the distribution of the
Walsh spectrum of    ternary 2-plateaued function $f$  is shown as follows.

i) For the case $n$ is odd,
\begin{eqnarray*}
\widehat{\chi}_f(a)=\left\{
  \begin{array}{ll}
    0, & \hbox{occurs $3^{n}-3^{n-2}$ times,} \\
    \eta(\lambda)i^n3^{\frac{n}{2}+1}, & \hbox{occurs $3^{n-3}$ times,} \\
    \eta(\lambda)i^n3^{\frac{n}{2}+1}\omega, & \hbox{occurs $3^{n-3}+\eta(\lambda)i^{n+1}3^{\frac{n-3}{2}}$ times,} \\
    \eta(\lambda)i^n3^{\frac{n}{2}+1}\omega^2, & \hbox{occurs $3^{n-3}-\eta(\lambda)i^{n+1}3^{\frac{n-3}{2}}$ times}.
  \end{array}
\right.
\end{eqnarray*}

i) For the case $n$ is even,
\begin{eqnarray*}
\widehat{\chi}_f(a)=\left\{
  \begin{array}{ll}
    0, & \hbox{occurs $3^{n}-3^{n-2}$ times,} \\
    -\eta(\lambda)i^n3^{\frac{n}{2}+1}, & \hbox{occurs $3^{n-3}-2\eta(\lambda)i^{n}3^{\frac{n}{2}-2}$ times,} \\
    -\eta(\lambda)i^n3^{\frac{n}{2}+1}\omega, & \hbox{occurs $3^{n-3}+\eta(\lambda)i^{n}3^{\frac{n}{2}-2}$ times,} \\
    -\eta(\lambda)i^n3^{\frac{n}{2}+1}\omega^2, & \hbox{occurs $3^{n-3}+\eta(\lambda)i^{n}3^{\frac{n}{2}-2}$ times}.
  \end{array}
\right.
\end{eqnarray*}
\end{description}}

\begin{proof}
The  proof is  similar  to that of Theorem \ref{t1} and is omitted.
\end{proof}
{\rem By Theorems \ref{t1} and \ref{t2}, one can conclude that  a ternary function given by adding the product of the product of two linear functions to known ternary bent function ${\rm Tr}^{k}_{1}(\lambda x^{3^k+1})$ or ${\rm Tr}^{n}_{1}(\lambda x^{2})$ must be among ternary bent, near-bent and 2-plateaued function. It is interesting to find  other ternary bent functions $f(x)$ such that $f(x)+{\rm Tr}_{1}^{n}(ux){\rm Tr}_{1}^{n}(vx)$ are also bent.}

{\exa Let  $n=4$ and let $\alpha$ be the generator of $\mathbb{F}_{3^4}^*$
 with $\alpha^4 - \alpha^3 - 1=0$.
\begin{description}
  \item[1)] Take $\lambda=\alpha$, $u=\alpha^4$ and $v=\alpha^7$. Then $f$ defined by (\ref{e12}) is
$$f(x)={\rm Tr}^{4}_{1}(\alpha x^{2})+{\rm Tr}_{1}^{4}(\alpha^4x){\rm Tr}_{1}^{4}(\alpha^7x).$$
By a Magma program, we can see that  ${\rm Tr}_{1}^{4}(\lambda^{-1}uv)=2,{\rm Tr}_{1}^{4}(\lambda^{-1}u^{2})=2, {\rm Tr}_{1}^{4}(\lambda^{-1}v^{2})=0$ and $f$ is bent.  This is consistent with the results given in Theorem \ref{t2}.
   \item[2)]Take $\lambda=\alpha$, $u=\alpha^4$ and $v=\alpha^8$. Then $f$ defined by (\ref{e12}) is
$$f(x)={\rm Tr}^{4}_{1}(\alpha x^{2})+{\rm Tr}_{1}^{4}(\alpha^4x){\rm Tr}_{1}^{4}(\alpha^8x).$$
By a Magma program, we can see that  ${\rm Tr}_{1}^{4}(\lambda^{-1}uv)=1,{\rm Tr}_{1}^{4}(\lambda^{-1}u^{2})=2, {\rm Tr}_{1}^{4}(\lambda^{-1}v^{2})=0$ and $f$ is near-bent.

  \item[3)] Take $\lambda=\alpha$, $u=\alpha^{16}$ and $v=\alpha^{8}$. Then $f$ defined by (\ref{e12}) is
$$f(x)={\rm Tr}^{4}_{1}(\alpha x^{2})+{\rm Tr}_{1}^{4}(\alpha^{16}x){\rm Tr}_{1}^{4}(\alpha^{8}x).$$
By a Magma program, we can see that  ${\rm Tr}_{1}^{4}(\lambda^{-1}uv)=1$, ${\rm Tr}_{1}^{4}(\lambda^{-1}u^{2})=0$, ${\rm Tr}_{1}^{4}(\lambda^{-1}v^{2})=0$ and $f$ is 2-plateaued. The the value distribution
of the Walsh transform of $f$ is
\begin{eqnarray*}
\widehat{\chi}_f(a)=\left\{
  \begin{array}{ll}
    0, & \hbox{occurs $72$ times,} \\
    27, & \hbox{occurs $5$ times,} \\
    27\omega, & \hbox{occurs $2$ times,} \\
    27\omega^2, & \hbox{occurs $2$ times.}
  \end{array}
\right.
\end{eqnarray*}
\end{description}
 Thus, our computer experiments are consistent with the results given in Theorem \ref{t2}.}

Below, we identify $\mathbb{F}_{3^{n}}$ (where $n=2k$) with $\mathbb{F}_{3^{k}}\times\mathbb{F}_{3^{k}}$.    For $a=(a_1,a_2), b=(b_1,b_2)\in\mathbb{F}_{3^{k}}\times\mathbb{F}_{3^{k}}$, the scalar product in $\mathbb{F}_{3^{n}}$ can be  defined as $\langle(a_1,a_2),(b_1,b_2) \rangle ={\rm Tr}^{k}_{1}(a_1b_1+a_2b_2).$
The well-known Maiorana-McFarland class of ternary bent functions  can be defined
as
$$g(x,y)={\rm Tr}^{k}_{1}(x\pi(y))+h(y), (x,y)\in\mathbb{F}_{3^{k}}\times\mathbb{F}_{3^{k}}$$
where  $\pi:\mathbb{F}_{3^{k}}\rightarrow\mathbb{F}_{3^{k}}$ is a permutation and $h$ is a ternary  function over $\mathbb{F}_{3^{k}}$, and its dual   is given by
$$g^*(x,y)=-{\rm Tr}^{k}_{1}(y\pi^{-1}(x))+h(\pi^{-1}(x))$$
where $\pi^{-1}$ denotes the inverse mapping of the permutation  $\pi$  \cite{kumar1985generalized}. This together with the definition of the dual function implies that for each $a=(a_1, a_2)\in\mathbb{F}_{3^{n}}$
\begin{align}\label{be21}
\widehat{\chi}_g(a_1,a_2)=3^k\omega^{-{\rm Tr}^{k}_{1}(a_2\pi^{-1}(a_1))+h(\pi^{-1}(a_1))}.
\end{align}

Similar to Lemma \ref{l5}, we have the following result.
{\lem \label{pl4} Let $n=2k$ be a even positive integer and $(u_1,u_2),  (v_1,v_2)\in \mathbb{F}_{3^{k}}\times\mathbb{F}_{3^{k}}$. Let $g(x,y)$ be a ternary function  defined on $\mathbb{F}_{3^{k}}\times\mathbb{F}_{3^{k}}$.  Let $f(x, y)$ be the ternary function defined as $$f(x, y)=g(x, y)+{\rm Tr}_{1}^{k}(u_1x+u_2y){\rm Tr}_{1}^{k}(v_1x+v_2y).$$ Then for any $(a_1,a_2)\in \mathbb{F}_{3^{k}}\times\mathbb{F}_{3^{k}}$, the corresponding Walsh transform coefficient of $f(x, y)$ is equal to
\begin{align*}
 \widehat{\chi}_f(a_1,a_2)=&\frac{1}{3}[\widehat{\chi}_g(a_1,a_2)+\widehat{\chi}_g(a_1+u_1,a_2+u_2)+ \widehat{\chi}_g(a_1-u_1,a_2-u_2)  \notag\\
+&\widehat{\chi}_g(a_1-v_1,a_2-v_2)+\omega\widehat{\chi}_g(a_1-v_1+u_1,a_2-v_2+u_2)
\notag\\+& \omega^2\widehat{\chi}_g(a_1-v_1-u_1,a_2-v_2-u_2)
+\widehat{\chi}_g(a_1+v_1,a_2+v_2)\notag\\+&\omega^2\widehat{\chi}_g(a_1+v_1+u_1,a_2+v_2+u_2)+ \omega\widehat{\chi}_g(a_1+v_1-u_1,a_2+v_2-u_2)]
\end{align*}
where $\omega$ is a primitive $3$-rd root of unity. }

 The following  theorem will employ the linearized permutation polynomial over $\mathbb{F}_{3^{k}}$ to give new ternary bent functions from   the class of Maiorana-McFarland.

{\thm\label{t3} Let $A$ be the subset of $\mathbb{F}_{3}^3$ defined as above. Let $n=2k$ and $u=(u_1,u_2),v=(v_1,v_2)$ be two  nonzero elements in $\mathbb{F}_{3^{k}}\times\mathbb{F}_{3^{k}}$. Assume that  $\pi$ is a linearized permutation polynomial over $\mathbb{F}_{3^{k}}$. Let $f(x,y)$ be the ternary function given by
\begin{align}\label{pb22}
f(x,y)={\rm Tr}^{k}_{1}(x\pi(y))+{\rm Tr}^{k}_{1}(y)+{\rm Tr}^{k}_{1}(u_1x+u_2y){\rm Tr}^{k}_{1}(v_1x+v_2y).
\end{align}

\begin{description}
  \item[1)] If  $\big({\rm Tr}^{k}_{1}(u_2\pi^{-1}(v_1)+v_2\pi^{-1}(u_1)),{\rm Tr}^{k}_{1}(u_2\pi^{-1}(u_1)),{\rm Tr}^{k}_{1}(v_2\pi^{-1}(v_1)) \big)\in
      \mathbb{F}_3^{3}\setminus (A\cup\{(2,0,0)\})$,
       then $f(x,y)$ is bent.
  \item[2)] If $\big({\rm Tr}^{k}_{1}(u_2\pi^{-1}(v_1)+v_2\pi^{-1}(u_1)),{\rm Tr}^{k}_{1}(u_2\pi^{-1}(u_1)),{\rm Tr}^{k}_{1}(v_2\pi^{-1}(v_1)) \big)\in
      A$, then $f(x,y)$ is near-bent.
\item[3)] If $\big({\rm Tr}^{k}_{1}(u_2\pi^{-1}(v_1)+v_2\pi^{-1}(u_1)),{\rm Tr}^{k}_{1}(u_2\pi^{-1}(u_1)),{\rm Tr}^{k}_{1}(v_2\pi^{-1}(v_1)) \big)=(2,0,0)$, then $f(x,y)$ is a ternary 2-plateaued function.

\end{description}}

\begin{proof}
Let $g(x,y)={\rm Tr}^{k}_{1}(x\pi(y))+{\rm Tr}^{k}_{1}(y)$. Applying Lemma \ref{pl4}, for each $(a_1,a_2)\in\mathbb{F}_{3^{k}}\times\mathbb{F}_{3^{k}}$, we have
\begin{align*}
 \widehat{\chi}_f(a_1,a_2)=\bigtriangleup_1+\bigtriangleup_2+\bigtriangleup_3,
\end{align*}
where
\begin{align*}
 \bigtriangleup_1=&\frac{1}{3}[\widehat{\chi}_g(a_1,a_2)+\widehat{\chi}_g(a_1+u_1,a_2+u_2)+ \widehat{\chi}_g(a_1-u_1,a_2-u_2)],
\end{align*}
\begin{align*}
 \bigtriangleup_2=&\frac{1}{3}[\widehat{\chi}_g(a_1-v_1,a_2-v_2)+\omega\widehat{\chi}_g(a_1-v_1+u_1,a_2-v_2+u_2)
 \\&+ \omega^2\widehat{\chi}_g(a_1-v_1-u_1,a_2-v_2-u_2)].
\end{align*}
and
\begin{align*}
 \bigtriangleup_3=&\frac{1}{3}[\widehat{\chi}_g(a_1+v_1,a_2+v_2)+\omega^2\widehat{\chi}_g(a_1+v_1+u_1,a_2+v_2+u_2)
 \\&+ \omega\widehat{\chi}_g(a_1+v_1-u_1,a_2+v_2-u_2)].
\end{align*}
Note that $\pi$ is a linearized permutation polynomial, and thus, $\pi^{-1}$ is also a linearized permutation polynomial. From (\ref{be21}), we can compute the sums $\bigtriangleup_1$, $\bigtriangleup_2$ and $\bigtriangleup_3$ respectively.
\begin{align}\label{p21}
 \bigtriangleup_1=&\frac{1}{3}\widehat{\chi}_g(a_1,a_2)[1+\omega^{-{\rm Tr}^{k}_{1}(a_2\pi^{-1}(u_1))-{\rm Tr}^{k}_{1}(u_2\pi^{-1}(a_1))+{\rm Tr}^{k}_{1}(\pi^{-1}(u_1))-{\rm Tr}^{k}_{1}(u_2\pi^{-1}(u_1))}
\notag\\&+\omega^{{\rm Tr}^{k}_{1}(a_2\pi^{-1}(u_1))+{\rm Tr}^{k}_{1}(u_2\pi^{-1}(a_1))-{\rm Tr}^{k}_{1}(\pi^{-1}(u_1))-{\rm Tr}^{k}_{1}(u_2\pi^{-1}(u_1))}]
\end{align}
Similarly, we have
\begin{align}\label{p22}
 \bigtriangleup_2=&\frac{1}{3}\widehat{\chi}_g(a_1,a_2)\omega^{{\rm Tr}^{k}_{1}(a_2\pi^{-1}(v_1))+{\rm Tr}^{k}_{1}(v_2\pi^{-1}(a_1))-{\rm Tr}^{k}_{1}(\pi^{-1}(v_1))-{\rm Tr}^{k}_{1}(v_2\pi^{-1}(v_1))}
\notag\\&\times[1+\omega^{1-{\rm Tr}^{k}_{1}(a_2\pi^{-1}(u_1))-{\rm Tr}^{k}_{1}(u_2\pi^{-1}(a_1))+{\rm Tr}^{k}_{1}(\pi^{-1}(u_1))-{\rm Tr}^{k}_{1}(u_2\pi^{-1}(u_1))}
\notag\\&\times\omega^{{\rm Tr}^{k}_{1}(u_2\pi^{-1}(v_1))+{\rm Tr}^{k}_{1}(v_2\pi^{-1}(u_1))}
\notag\\&+\omega^{2+{\rm Tr}^{k}_{1}(a_2\pi^{-1}(u_1))+{\rm Tr}^{k}_{1}(u_2\pi^{-1}(a_1))-{\rm Tr}^{k}_{1}(\pi^{-1}(u_1))-{\rm Tr}^{k}_{1}(u_2\pi^{-1}(u_1))}
\notag\\&\times\omega^{-{\rm Tr}^{k}_{1}(u_2\pi^{-1}(v_1))-{\rm Tr}^{k}_{1}(v_2\pi^{-1}(u_1))}]
\end{align}
and
\begin{align}\label{p23}
 \bigtriangleup_3=&\frac{1}{3}\widehat{\chi}_g(a_1,a_2)\omega^{-{\rm Tr}^{k}_{1}(a_2\pi^{-1}(v_1))-{\rm Tr}^{k}_{1}(v_2\pi^{-1}(a_1))+{\rm Tr}^{k}_{1}(\pi^{-1}(v_1))-{\rm Tr}^{k}_{1}(v_2\pi^{-1}(v_1))}
\notag\\&\times[1+\omega^{2-{\rm Tr}^{k}_{1}(a_2\pi^{-1}(u_1))-{\rm Tr}^{k}_{1}(u_2\pi^{-1}(a_1))+{\rm Tr}^{k}_{1}(\pi^{-1}(u_1))-{\rm Tr}^{k}_{1}(u_2\pi^{-1}(u_1))}
\notag\\&\times\omega^{-{\rm Tr}^{k}_{1}(u_2\pi^{-1}(v_1))-{\rm Tr}^{k}_{1}(v_2\pi^{-1}(u_1))}
\notag\\&+\omega^{1+{\rm Tr}^{k}_{1}(a_2\pi^{-1}(u_1))+{\rm Tr}^{k}_{1}(u_2\pi^{-1}(a_1))-{\rm Tr}^{k}_{1}(\pi^{-1}(u_1))-{\rm Tr}^{k}_{1}(u_2\pi^{-1}(u_1))}
\notag\\&\times\omega^{{\rm Tr}^{k}_{1}(u_2\pi^{-1}(v_1))+{\rm Tr}^{k}_{1}(v_2\pi^{-1}(u_1))}
].
\end{align}
For convenience of presentation, we let $t_0={\rm Tr}^{k}_{1}(u_2\pi^{-1}(v_1))+{\rm Tr}^{k}_{1}(v_2\pi^{-1}(u_1))$, $t_1={\rm Tr}^{k}_{1}(u_2\pi^{-1}(u_1))$ and $t_2={\rm Tr}^{k}_{1}(v_2\pi^{-1}(v_1))$ and denote $c_1={\rm Tr}^{k}_{1}(a_2\pi^{-1}(u_1))+{\rm Tr}^{k}_{1}(u_2\pi^{-1}(a_1))-{\rm Tr}^{k}_{1}(\pi^{-1}(u_1))$ and $c_2={\rm Tr}^{k}_{1}(a_2\pi^{-1}(v_1))+{\rm Tr}^{k}_{1}(v_2\pi^{-1}(a_1))-{\rm Tr}^{k}_{1}(\pi^{-1}(v_1))$.  Thus,  (\ref{p21}), (\ref{p22}) and (\ref{p23}) can be written as

\begin{eqnarray*}
\bigtriangleup_1=&\frac{1}{3}\widehat{\chi}_g(a_1,a_2)\left( 1+\omega^{-c_1-t_1}+\omega^{c_1-t_1}\right),
\end{eqnarray*}
\begin{align*}
\bigtriangleup_2=&\frac{1}{3}\widehat{\chi}_g(a_1,a_2))\omega^{c_2-t_2}
( 1+\omega^{1-c_1+t_0-t_1}
+\omega^{2+c_1-t_0-t_1})
\end{align*}
and
\begin{align*}
\bigtriangleup_3=&\frac{1}{3}\widehat{\chi}_g(a_1,a_2))\omega^{-c_2-t_2}
( 1+\omega^{2-c_1-t_0-t_1}
+\omega^{1+c_1+t_0-t_1}),
\end{align*}
Similar as the proof of Theorem \ref{t1}, detailed discussing of the possible values of $t_i$ and $c_i$ leads to the desired conclusion.
\end{proof}

{\rem To obtain our constructions in Theorem \ref{t3}, we need to know the compositional inverse of a given linearized permutation polynomial
over $\mathbb{F}_{3^k}$.
 It is clear that the simplest suitable linearized permutation polynomial $\pi$ over $\mathbb{F}_{3^k}$ in Theorem \ref{t3} is $y^{3^i}$
where
$0 \leq i \leq n-1 $. In addition, the linearized permutation binomials in Lemma \ref{pppl1} with $p=3$ can also be employed to construct ternary functions in Theorem \ref{t3}.}

{\exa  Let $k=4$ and $n=8$.  Let $\alpha$ be the generator of $\mathbb{F}_{3^4}^*$ with  $\alpha^4 - \alpha^3 - 1=0$. Take $\pi(y)=y^9+\alpha y$, by Lemma \ref{pppl1}, we can get  $\pi^{-1}(y)=(\alpha^{10}-1)^{-1}(\alpha^9y- y^9)$.
\begin{description}
  \item[1)] Let $u=(u_1, u_2)=(\alpha^{4},\alpha^{5})$  and $v=(v_1, v_2)=(\alpha^{10},\alpha^2)$. Then  $f(x,y)$ defined by (\ref{pb22}) is
$$f(x,y)={\rm Tr}^{4}_{1}(x(y^9+\alpha y))+{\rm Tr}_{1}^{4}(y)+{\rm Tr}^{4}_{1}(\alpha^{4}x+\alpha^{5}y){\rm Tr}^{4}_{1}(\alpha^{10}x+\alpha^2y).$$
Using Magma, we can verify that ${\rm Tr}^{4}_{1}(u_2\pi^{-1}(v_1))+{\rm Tr}^{k}_{1}(v_2\pi^{-1}(u_1))=1$,  ${\rm Tr}^{4}_{1}(u_2\pi^{-1}(u_1)))=0$, ${\rm Tr}^{4}_{1}(v_2\pi^{-1}(v_1)))=0$ and $f(x,y)$ is bent.
 \item[2)] Let $u=(u_1, u_2)=(\alpha^{10},\alpha^{11})$  and $v=(v_1, v_2)=(\alpha^{10},\alpha^{73})$. Then  $f(x,y)$ defined by (\ref{pb22}) is
$$f(x,y)={\rm Tr}^{4}_{1}(x(y^9+\alpha y))+{\rm Tr}_{1}^{4}(y)+{\rm Tr}^{4}_{1}(\alpha^{10}x+\alpha^{11}y){\rm Tr}^{4}_{1}(\alpha^{10}x+\alpha^{73}y).$$
Using Magma, we can verify that ${\rm Tr}^{4}_{1}(u_2\pi^{-1}(v_1))+{\rm Tr}^{4}_{1}(v_2\pi^{-1}(u_1))=2$,  ${\rm Tr}^{4}_{1}(u_2\pi^{-1}(u_1)))=2$, ${\rm Tr}^{4}_{1}(v_2\pi^{-1}(v_1)))=0$ and $f(x,y)$ is near-bent.

  \item[3)] Let $u=(u_1, u_2)=(\alpha^{4},\alpha^{5})$  and $v=(v_1, v_2)=(\alpha^{10},\alpha^{46})$. Then  $f(x,y)$ defined by (\ref{pb22}) is
$$f(x,y)={\rm Tr}^{4}_{1}(x(y^9+\alpha y))+{\rm Tr}_{1}^{4}(y)+{\rm Tr}^{4}_{1}(\alpha^{4}x+\alpha^{5}y){\rm Tr}^{4}_{1}(\alpha^{10}x+\alpha^{46}y).$$
Using Magma, we can verify that ${\rm Tr}^{4}_{1}(u_2\pi^{-1}(v_1))+{\rm Tr}^{4}_{1}(v_2\pi^{-1}(u_1))=2$,  ${\rm Tr}^{4}_{1}(u_2\pi^{-1}(u_1)))=0$, ${\rm Tr}^{4}_{1}(v_2\pi^{-1}(v_1)))=0$ and $f(x,y)$ is 2-plateaued.
\end{description}
Our computer experiments are consistent with the results given in Theorem \ref{t3}.}

\section{ Two new class of non-quadratic $p$-ary bent functions of algebraic degree $p$ }

In \cite{xuguangkui}, the authors provided a class of non-quadratic $p$-ary bent functions $f(x)={\rm Tr}^{k}_{1}(\lambda x^{p^i+1})+c{\rm Tr}_{1}^{n}(x)^{p-1}$ of the algebraic degree $p-1$, where $\frac{n}{\gcd(i,n)} $ is odd.  In this section,  two  new class of non-quadratic $p$-ary bent functions of  algebraic degree $p$  will be obtained. Before doing this, we need the following lemma.
{\lem \label{l7} Let  $p$ be an odd prime and $n$ be a positive  integer.  Let $g(x)$ be a $p$-ary function defined on $\mathbb{F}_{p^{n}}$. Let $f(x)$ be the $p$-ary function defined by
$$f(x)=g(x)-{\rm Tr}_{1}^{n}(ux){\rm Tr}_{1}^{n}(x)^{p-1},$$ where $c\in\mathbb{F}_{p}^*$ and $u\in\mathbb{F}_{p^n}^*$ . Then, for  every $a\in \mathbb{F}_{p^{n}}$  the  corresponding Walsh transform coefficient of $f(x)$ is equal to
\begin{eqnarray*}
 \widehat{\chi}_f(a)&=&\frac{1}{p}\sum_{j=0}^{p-1}\widehat{\chi}_g(a+j)-\frac{1}{p}\sum_{j=0}^{p-1}\widehat{\chi}_g(a+u+j)+\widehat{\chi}_g(a+u),
\end{eqnarray*}}
where $\omega$ is a primitive $p$-th root of unity.

\begin{proof}
For $i=0,1,\cdots,p-1$, we define  $T_i=\{x\in\mathbb{F}_{p^{n}}|{\rm Tr}_{1}^{n}(x)=i\}$. To calculate the Walsh transform coefficient of $f(x)$ evaluated at $a\in\mathbb{F}_{p^{n}}$,  we define the following exponential sums

 $$S_i(a)=\sum_{x\in T_i}\omega^{g(x)-{\rm Tr}_{1}^{n}(ax)}, Q_i(a+u)=\sum_{x\in T_i}\omega^{g(x)-{\rm Tr}_{1}^{n}((a+u)x)},$$
 where $i=0,1,\cdots,p-1$.
Then we have
\begin{align*}
&\widehat{\chi}_f(a)
\\=&\sum_{x\in\mathbb{F}_{p^{n}}}\omega^{f(x)-{\rm Tr}_{1}^{n}(ax)}
\\=&\sum_{x\in\mathbb{F}_{p^n}}\omega^{g(x)-{\rm Tr}_{1}^{n}(ux){\rm Tr}_{1}^{n}(x)^{p-1}-{\rm Tr}_{1}^{n}(ax)}
\\=&\sum_{x\in T_0}\omega^{g(x)-{\rm Tr}_{1}^{n}(ax)}+\sum_{x\in T_1}\omega^{g(x)-{\rm Tr}_{1}^{n}((a+u)x)}+\cdots+\sum_{x\in T_{p-1}}\omega^{g(x)-{\rm Tr}_{1}^{n}((a+u)x)}
\\=&S_0(a)-Q_0(a+u)+\widehat{\chi}_g(a+u).
\end{align*}
 For a  $j\in \mathbb{F}_{p}$, we can derive
\begin{eqnarray}\label{epp21}
\sum_{x\in T_i}\omega^{g(x)-{\rm Tr}_{1}^{n}((a+j)x)}=\sum_{x\in T_i}\omega^{g(x)-{\rm Tr}_{1}^{n}(ax)-j{\rm Tr}_{1}^{n}(x)}=\omega^{-ji}S_i(a).
\end{eqnarray}
It then follows that
$\widehat{\chi}_g(a+j)=\sum_{i=0}^{p-1}\omega^{-ji}S_i(a).$ For $j\in\{0,1,\cdots, p-1\}$,  we can get
\begin{eqnarray}\label{eg21}
\left\{
  \begin{array}{ll}
    \widehat{\chi}_g(a)=\sum_{i=0}^{p-1}S_i(a) & \hbox{} \\
    \widehat{\chi}_g(a+1)=\sum_{i=0}^{p-1}\omega^{-i}S_i(a) & \hbox{} \\
    \cdots & \hbox{} \\
    \widehat{\chi}_g(a+p-1)=\sum_{i=0}^{p-1}\omega^{-(p-1)i}S_i(a) & \hbox{.}
  \end{array}
\right.
\end{eqnarray}
Note that $1+\omega+\cdots+\omega^{p-1}=0$. Adding  $p$ equations of (\ref{eg21}) gives
$$S_0(a)=\frac{1}{p}\sum_{j=0}^{p-1}\widehat{\chi}_g(a+j).$$
Substituting $a$ by $a + u$ in (\ref{epp21}) and repeating above process, we have
$$Q_0(a+u)=\frac{1}{p}\sum_{j=0}^{p-1}\widehat{\chi}_g(a+u+j).$$
Finally, we have
\begin{eqnarray*}
 \widehat{\chi}_f(a)&=&\frac{1}{p}\sum_{j=0}^{p-1}\widehat{\chi}_g(a+j)-\frac{1}{p}\sum_{j=0}^{p-1}\widehat{\chi}_g(a+u+j)+\widehat{\chi}_g(a+u).
\end{eqnarray*}
\end{proof}

{\thm \label{t5}   Let $n=2k$ with $k>1$ and
$\lambda \in \mathbb{F}_{p^{k}}^*$, $u\in \mathbb{F}_{p^{n}}\setminus\mathbb{F}_{p}$ such that ${\rm Tr}^{k}_{1}(\lambda^{-1})=0$ and ${\rm Tr}^{n}_{1}(\lambda^{-1}u)=0$. Then  the $p$-ary function
\begin{eqnarray}\label{ej18}
f(x)={\rm Tr}^{k}_{1}(\lambda x^{p^k+1})-{\rm Tr}_{1}^{n}(ux){\rm Tr}_{1}^{n}(x)^{p-1}
\end{eqnarray}
is a weakly regular bent function.   Moreover, for $a\in \mathbb{F}_{p^{n}}$ the corresponding Walsh transform coefficient of $f(x)$ is equal to
\begin{eqnarray*}
\widehat{\chi}_f(a)=\left\{
  \begin{array}{ll}
    -p^k\omega^{-{\rm Tr}^{k}_{1}(\lambda^{-1}a^{p^k+1})}, & \hbox{if ${\rm Tr}^{n}_{1}(\lambda^{-1}a)=0$ ;} \\
     -p^k\omega^{-{\rm Tr}^{k}_{1}(\lambda^{-1}(a+u)^{p^k+1})}, & \hbox{if ${\rm Tr}^{n}_{1}(\lambda^{-1}a)\neq0$.}
  \end{array}
\right.
\end{eqnarray*} }
\begin{proof}
Let $g(x)={\rm Tr}^{k}_{1}(\lambda x^{p^k+1})$. By lemma \ref{pl1}, $g(x)$ is a $p$-ary bent function and $$\widehat{\chi}_g(a)=-p^k\omega^{-{\rm Tr}^{k}_{1}(\lambda^{-1}a^{p^k+1})}$$ for $a\in \mathbb{F}_{p^{n}}$. It then follows from Lemma \ref{l7} that
\begin{align}\label{pa-24}
 \widehat{\chi}_f(a)=&\frac{1}{p}\sum_{j=0}^{p-1}\widehat{\chi}_g(a+j)-\frac{1}{p}\sum_{j=0}^{p-1}\widehat{\chi}_g(a+u+j)+\widehat{\chi}_g(a+u)
\notag\\=&-p^k\bigg(\frac{1}{p}\sum_{j=0}^{p-1}\omega^{-{\rm Tr}^{k}_{1}(\lambda^{-1}a^{p^k+1})-j{\rm Tr}^{k}_{1}(\lambda^{-1}(a^{p^k}+a))-j^2{\rm Tr}^{k}_{1}(\lambda^{-1})}
\notag\\&-\frac{1}{p}\sum_{j=0}^{p-1}\omega^{-{\rm Tr}^{k}_{1}(\lambda^{-1}(a+u)^{p^k+1})-j{\rm Tr}^{k}_{1}(\lambda^{-1}((a+u)^{p^k}+a+u))-j^2{\rm Tr}^{k}_{1}(\lambda^{-1})}
\notag\\&+\omega^{-{\rm Tr}^{k}_{1}(\lambda^{-1}(a+u)^{p^k+1})}\bigg)
\notag\\=&-p^k\bigg(\frac{1}{p}\sum_{j=0}^{p-1}\omega^{-{\rm Tr}^{k}_{1}(\lambda^{-1}a^{p^k+1})-j{\rm Tr}^{n}_{1}(\lambda^{-1}a)-j^2{\rm Tr}^{k}_{1}(\lambda^{-1})}
\notag\\&-\frac{1}{p}\sum_{j=0}^{p-1}\omega^{-{\rm Tr}^{k}_{1}(\lambda^{-1}(a+u)^{p^k+1})-j{\rm Tr}^{n}_{1}(\lambda^{-1}a)-j{\rm Tr}^{n}_{1}(\lambda^{-1}u)-j^2{\rm Tr}^{k}_{1}(\lambda^{-1})}
\notag\\&+\omega^{-{\rm Tr}^{k}_{1}(\lambda^{-1}(a+u)^{p^k+1})}\bigg)
\end{align}
Since ${\rm Tr}^{k}_{1}(\lambda^{-1})=0$ and ${\rm Tr}^{n}_{1}(\lambda^{-1}u)=0$, (\ref{pa-24})  can be simplified to
\begin{align*}
 \widehat{\chi}_f(a)=&-p^k\bigg(\omega^{-{\rm Tr}^{k}_{1}(\lambda^{-1}a^{p^k+1})}\frac{1}{p}\sum_{j=0}^{p-1}\omega^{-j{\rm Tr}^{n}_{1}(\lambda^{-1}a)}
\\&-\omega^{-{\rm Tr}^{k}_{1}(\lambda^{-1}(a+u)^{p^k+1})}\big(\frac{1}{p}\sum_{j=0}^{p-1}\omega^{-j{\rm Tr}^{n}_{1}(\lambda^{-1}a)}-1\big)\bigg)
\end{align*}
Clearly, if ${\rm Tr}^{n}_{1}(\lambda^{-1}a)=0$, then $$\widehat{\chi}_f(a)=-p^k\omega^{-{\rm Tr}^{k}_{1}(\lambda^{-1}a^{p^k+1})}.$$ If ${\rm Tr}^{n}_{1}(\lambda^{-1}a)\neq0$,
then $$\widehat{\chi}_f(a)=-p^k\omega^{-{\rm Tr}^{k}_{1}(\lambda^{-1}(a+u)^{p^k+1})}.$$
Therefore $|\widehat{\chi}_f(a)|=p^k$ for all $a\in \mathbb{F}_{p^{n}}$.
\end{proof}

{\rem  1) If $k=1$, i.e., $n=2$, then ${\rm Tr}^{k}_{1}(\lambda^{-1})=\lambda^{-1}\neq0$ for $\lambda \in\mathbb{F}_{p}^*$.  It is clear that ${\rm Tr}_{1}^{n}(ux){\rm Tr}_{1}^{n}(x)^{p-1}=u{\rm Tr}_{1}^{n}(x)$ if $u\in\mathbb{F}_{p}^*$.

2) It was shown  in \cite{hou2004p} that the  algebraic degree of  $p$-ary weakly bent functions with $(p-1)n\geq 4$ over $\mathbb{F}_{p^{n}}$ is at most  $\frac{(p-1)n}{2}$. Clearly,  the algebraic degree of  ${\rm Tr}_{1}^{n}(ux){\rm Tr}_{1}^{n}(x)^{p-1}$ is at most  $p$.  Expanding ${\rm Tr}_{1}^{n}(ux){\rm Tr}_{1}^{n}(x)^{p-1}$ by  the definition of trace function,   we can verify that there is a term $(u^p-u)x^{p+(p-1)}$ in $f(x)$. Note that $u^p-u\neq0$ beause $u\in \mathbb{F}_{p^{n}}\setminus\mathbb{F}_{p}$. This implies that the algebraic degree  of the $p$-ary bent function defined by (\ref{ej18}) is equal to $p$.
  Up to the authors' knowledge, this is the first class of $p$-ary bent functions of algebraic degree $p$. In other words, we proved that for any odd prime number $p$, there is a bent function of algebraic degree $p$. For an odd prime $p$, some known classes of $p$-ary monomial and binomial bent functions were
listed in Table II \cite{jia2012class} and Table 1 \cite{zheng} respectively. Note that algebraic degree is an affine invariant. According to Table II \cite{jia2012class} and Table 1 \cite{zheng}, we conclude that there exist bent functions of  the
form as in (\ref{ej18}) which are affinely inequivalent to all known ones
listed in Table II \cite{jia2012class} and Table 1 \cite{zheng}.}

{\exa Let $p=7$, $k=2$ and $n=4$.   Let $\alpha$ be the generator of $\mathbb{F}_{5^6}^*$
 with $\alpha^4+5\alpha^2 +4\alpha+3=0$. Take $\lambda=\alpha^{200}$, $u=\alpha^{90}$. Then $f$ defined by (\ref{ej18}) is
$$f(x)={\rm Tr}^{2}_{1}(\alpha^{200} x^{50})-{\rm Tr}_{1}^{4}(\alpha^{90}x){\rm Tr}_{1}^{4}(x)^6.$$
Using Magma, we can verify  that  ${\rm Tr}_{1}^{2}(\lambda^{-1})=0$,  ${\rm Tr}_{1}^{4}(\lambda^{-1}u)=0$ and $f$ is bent,   which is consistent with the results given in Theorem \ref{t5}.}

Similar to above, by Lemmas \ref{l3} and \ref{l7}, we can obtain another calss of $p$-ary bent functions of algebraic degree $p$.
{\thm\label{t6}    For any positive integer $n>2$,  let  $\lambda\in \mathbb{F}_{p^{n}}^*$ and $u\in \mathbb{F}_{p^{n}}\setminus\mathbb{F}_{p}$ . Then the $p$-ary  function $f$ defined as
\begin{eqnarray}\label{ej19}
f(x)={\rm Tr}^{n}_{1}(\lambda x^{2})-{\rm Tr}_{1}^{n}(ux){\rm Tr}_{1}^{n}(x)^{p-1}
\end{eqnarray}
is a weakly regular bent function if ${\rm Tr}^{n}_{1}\left(\frac{1}{4\lambda}\right)=0$ and ${\rm Tr}^{n}_{1}\left(\frac{u}{4\lambda}\right)=0$. }
{\rem   When $n=2$, let   $\alpha$ be the generator of $\mathbb{F}_{p^{2}}^*$.
The equation ${\rm Tr}^{2}_{1}\left(\frac{1}{4x}\right)=\frac{1}{4x}+(\frac{1}{4x})^p=0$, i.e., $x^{p-1}=-1$
has $p-1$ solutions   $\alpha^{\frac{(p+1)i}{2}}$ in $\mathbb{F}_{p^{2}}^*$, where $ i=2t-1$ for $1\leq t\leq p-1 $. For any  solution $\lambda$ of $x^{p-1}=-1$, the equation  ${\rm Tr}^{n}_{1}\left(\frac{y}{4\lambda}\right)=0$, i.e., $\frac{y}{4\lambda}+(\frac{y}{4\lambda})^p
=\frac{y}{4\lambda}+\frac{1}{4\lambda}(\frac{1}{\lambda})^{p-1}y^p=\frac{y}{4\lambda}-\frac{1}{4\lambda}y^p=0$ leads to   $y\in \mathbb{F}_{p}$, which is contradict with our assumption $u\in \mathbb{F}_{p^{n}}\setminus\mathbb{F}_{p}$. Therefore we assume that $n>2$ in Theorem \ref{t6}.}

{\exa  Let $p=5$ and $n=3$.   Let $\alpha$ be the generator of $\mathbb{F}_{5^3}^*$
 with $\alpha^3 +3\alpha+3=0$. Take $\lambda=\alpha^{9}$, $u=\alpha^{14}$. Then $f$ defined by (\ref{ej19}) is
$$f(x)={\rm Tr}^{3}_{1}(\alpha^{9} x^{2})-{\rm Tr}_{1}^{3}(\alpha^{14}x){\rm Tr}_{1}^{3}(x)^4.$$
Using Magma, we can verify  that  ${\rm Tr}_{1}^{3}((4\lambda)^{-1})=0$,  ${\rm Tr}_{1}^{3}((4\lambda)^{-1}u)=0$ and $f$ is bent,   which is consistent with the results given in Theorem \ref{t6}.}

\section{Conclusion}
The main purpose of this paper is to provide  constructions of  $p$-ary  functions with low Walsh spectra.
We succeed in  constructing  more  quadratic ternary bent, near-bent and   2-plateaued functions, and determining the distribution of the Walsh
spectrum of some 2-plateaued functions constructed in this paper. In addition, two classes of non-quadratic $p$-ary bent functions are obtained. Furthermore, our computer experiments show that there are more $p$-ary bent functions of the form  (\ref{ej4}) which may be obtained via  $p$-ary Kasami and Sidelnikov monomials, such as $f_1(x)={\rm Tr}^{4}_{1}(\xi^6 x^{2})+2{\rm Tr}_{1}^{4}(x)^2$, $f_2(x)={\rm Tr}^{4}_{1}(\xi x^{26})+2{\rm Tr}_{1}^{4}(x)^2$ over $\mathbb{F}_{5^4}$, where $\xi$ is a primitive element of $\mathbb{F}_{5^4}$. Unfortunately, the bentness of theses functions can not be characterized by our method.  We leave this for future work.

\section*{Acknowledgement}

The work of this paper was supported by the National Natural Science Foundation of China (Grants 11371011 and 61403157), the Natural Science Foundation of the Anhui Higher Education
Institutions of China (Grants KJ2013B256 and  KJ2013B261) and the Natural Science Foundation of
 the Huainan Normal University  (Grant 2013XJ67).

\section{ Appendix}\label{ap1}
\noindent{\bf Computation of the value distribution of the Walsh transform of 2-plateaued function $f$ given by (\ref{ej12}).}

The condition that  ${\rm Tr}^{k}_{1}(\lambda^{-1}u^{3^k+1})={\rm Tr}^{k}_{1}(\lambda^{-1}v^{3^k+1})=0$ and ${\rm Tr}^{n}_{1}(\lambda^{-1}u^{3^k}v)=2$ leads to $u\neq\pm v$. Otherwise, assume that $u=\pm v$, then ${\rm Tr}^{n}_{1}(\lambda^{-1}u^{3^k}v)=\pm{\rm Tr}^{n}_{1}(\lambda^{-1}u^{3^k+1})=\pm{\rm Tr}^{k}_{1}({\rm Tr}^{n}_{k}(\lambda^{-1}u^{3^k+1}))=\pm 2{\rm Tr}^{k}_{1}(\lambda^{-1}u^{3^k+1})=2$, and thus, ${\rm Tr}^{k}_{1}(\lambda^{-1}u^{3^k+1})=\pm1$. This is contrary to the assumption that ${\rm Tr}^{k}_{1}(\lambda^{-1}u^{3^k+1})=0$.

 Let ${\rm Tr}^{k}_{1}(\lambda^{-1}a^{3^k+1})=c_0$ and  denote by $N_i$ the number of $a\in\mathbb{F}_{3^{n}}$ such that $\widehat{\chi}_f(a)= -3^{k+1}\omega^{i}$, where $i =0, 1, 2.$
The equality  (\ref{pp13}) implies that if ${\rm Tr}^{k}_{1}(\lambda^{-1}a^{3^k+1})=0$, ${\rm Tr}^{n}_{1}(\lambda^{-1}a^{3^k}u)=0$ and ${\rm Tr}^{n}_{1}(\lambda^{-1}a^{3^k}v)=0$, then $\widehat{\chi}_f(a)=-3^{k+1}$. Hence we have
\begin{align}\label{pp14}
N_{0}=&\frac{1}{27}\sum_{a\in\mathbb{F}_{3^n}}\sum_{x\in\mathbb{F}_{3}}\omega^{x{\rm Tr}^{k}_{1}(\lambda^{-1}a^{3^k+1})}\sum_{y\in\mathbb{F}_{3}}\omega^{y{\rm Tr}^{n}_{1}(\lambda^{-1}a^{3^k}u)}\sum_{z\in\mathbb{F}_{3}}\omega^{z{\rm Tr}^{n}_{1}(\lambda^{-1}a^{3^k}v)}
\notag\\=&\frac{1}{27}\sum_{a\in\mathbb{F}_{3^n}}(1+\omega^{{\rm Tr}^{k}_{1}(\lambda^{-1}a^{3^k+1})}+\omega^{-{\rm Tr}^{k}_{1}(\lambda^{-1}a^{3^k+1})})
(1+\omega^{{\rm Tr}^{n}_{1}(\lambda^{-1}a^{3^k}u)}
\notag\\&+\omega^{-{\rm Tr}^{n}_{1}(\lambda^{-1}a^{3^k}u)})
(1+\omega^{{\rm Tr}^{n}_{1}(\lambda^{-1}a^{3^k}v)}+\omega^{-{\rm Tr}^{n}_{1}(\lambda^{-1}a^{3^k}v)})
\notag\\=&\frac{1}{27}\sum_{a\in\mathbb{F}_{3^n}}
(1+\omega^{{\rm Tr}^{k}_{1}(\lambda^{-1}a^{3^k+1})}+\omega^{-{\rm Tr}^{k}_{1}(\lambda^{-1}a^{3^k+1})})
\big(1+\omega^{{\rm Tr}^{n}_{1}(\lambda^{-1}a^{3^k}v)}
\notag\\&+\omega^{-{\rm Tr}^{n}_{1}(\lambda^{-1}a^{3^k}v)}
+\omega^{{\rm Tr}^{n}_{1}(\lambda^{-1}a^{3^k}u)}+\omega^{{\rm Tr}^{n}_{1}(\lambda^{-1}a^{3^k}u)+{\rm Tr}^{n}_{1}(\lambda^{-1}a^{3^k}v)}
\notag\\&+\omega^{{\rm Tr}^{n}_{1}(\lambda^{-1}a^{3^k}u)-{\rm Tr}^{n}_{1}(\lambda^{-1}a^{3^k}v)}
+\omega^{-{\rm Tr}^{n}_{1}(\lambda^{-1}a^{3^k}u)}+\omega^{-{\rm Tr}^{n}_{1}(\lambda^{-1}a^{3^k}u)+{\rm Tr}^{n}_{1}(\lambda^{-1}a^{3^k}v)}
\notag\\&+\omega^{-{\rm Tr}^{n}_{1}(\lambda^{-1}a^{3^k}u)-{\rm Tr}^{n}_{1}(\lambda^{-1}a^{3^k}v)}\big).
\end{align}
Recalling $u, v \in \mathbb{F}_{3^{n}}^* $ and $u\neq\pm v$, we have
 \begin{align}\label{pp15}
\sum_{a\in\mathbb{F}_{3^n}}\omega^{\pm{\rm Tr}^{n}_{1}(\lambda^{-1}a^{3^k}u)}&=\sum_{a\in\mathbb{F}_{3^n}}\omega^{\pm{\rm Tr}^{n}_{1}(\lambda^{-1}a^{3^k}v)}=\sum_{a\in\mathbb{F}_{3^n}}\omega^{-{\rm Tr}^{n}_{1}(\lambda^{-1}a^{3^k}(u\pm v))}
\notag\\&=\sum_{a\in\mathbb{F}_{3^n}}\omega^{-{\rm Tr}^{n}_{1}(\lambda^{-1}a^{3^k}( v\pm u))}=0.
\end{align}
From (\ref{pp15}), the equality (\ref{pp14}) can be simplified to
\begin{align}\label{pp16}
N_{0}=&3^{n-3}+\frac{1}{27}\sum_{a\in\mathbb{F}_{3^n}}
(\omega^{{\rm Tr}^{k}_{1}(\lambda^{-1}a^{3^k+1})}+\omega^{-{\rm Tr}^{k}_{1}(\lambda^{-1}a^{3^k+1})})
\big(1+\omega^{{\rm Tr}^{n}_{1}(\lambda^{-1}a^{3^k}v)}
\notag\\&+\omega^{-{\rm Tr}^{n}_{1}(\lambda^{-1}a^{3^k}v)}
+\omega^{{\rm Tr}^{n}_{1}(\lambda^{-1}a^{3^k}u)}
+\omega^{{\rm Tr}^{n}_{1}(\lambda^{-1}a^{3^k}u)+{\rm Tr}^{n}_{1}(\lambda^{-1}a^{3^k}v)}
\notag\\&+\omega^{{\rm Tr}^{n}_{1}(\lambda^{-1}a^{3^k}u)-{\rm Tr}^{n}_{1}(\lambda^{-1}a^{3^k}v)}
+\omega^{-{\rm Tr}^{n}_{1}(\lambda^{-1}a^{3^k}u)}+\omega^{-{\rm Tr}^{n}_{1}(\lambda^{-1}a^{3^k}u)+{\rm Tr}^{n}_{1}(\lambda^{-1}a^{3^k}v)}
\notag\\&+\omega^{-{\rm Tr}^{n}_{1}(\lambda^{-1}a^{3^k}u)-{\rm Tr}^{n}_{1}(\lambda^{-1}a^{3^k}v)}\big).
\end{align}

Since raising elements of $\mathbb{F}_{3^n}^*$ to the power of $3^k + 1$ is a $3^k+1$ to 1 mapping on to $\mathbb{F}_{3^k}^*$,
\begin{align}\label{pp17}
\sum_{a\in\mathbb{F}_{3^n}}
\omega^{{\rm Tr}^{k}_{1}(\lambda^{-1}a^{3^k+1})}&=\sum_{a\in\mathbb{F}_{3^n}}
\omega^{-{\rm Tr}^{k}_{1}(\lambda^{-1}a^{3^k+1})}
\notag\\&=1+(3^k+1)\sum_{b\in\mathbb{F}_{3^k}^*}\omega^{{\rm Tr}^{k}_{1}(\lambda^{-1}b)}=-3^k
\end{align}
because $\lambda\in\mathbb{F}_{3^k}^*$.

From ${\rm Tr}^{k}_{1}(\lambda^{-1}u^{3^k+1})=0$, we obtain
\begin{align}
&\sum_{a\in\mathbb{F}_{3^n}}\omega^{{\rm Tr}^{k}_{1}(\lambda^{-1}a^{3^k+1})\pm{\rm Tr}^{n}_{1}(\lambda^{-1}a^{3^k}u)}
\notag\\&=\sum_{a\in\mathbb{F}_{3^n}}\omega^{{\rm Tr}^{k}_{1}(\lambda^{-1}a^{3^k+1})\pm{\rm Tr}^{n}_{1}(\lambda^{-1}a^{3^k}u)+{\rm Tr}^{k}_{1}(\lambda^{-1}u^{3^k+1})}
\notag\\&=\sum_{a\in\mathbb{F}_{3^n}}\omega^{{\rm Tr}^{k}_{1}(\lambda^{-1}(a\pm u)^{3^k+1})}=-3^k
\end{align}
and
\begin{align}
&\sum_{a\in\mathbb{F}_{3^n}}\omega^{-{\rm Tr}^{k}_{1}(\lambda^{-1}a^{3^k+1})\pm{\rm Tr}^{n}_{1}(\lambda^{-1}a^{3^k}u)}=-3^k.
\end{align}
Similarly, it follows from   ${\rm Tr}^{k}_{1}(\lambda^{-1}v^{3^k+1})=0$ that
\begin{align}
&\sum_{a\in\mathbb{F}_{3^n}}\omega^{{\rm Tr}^{k}_{1}(\lambda^{-1}a^{3^k+1})\pm{\rm Tr}^{n}_{1}(\lambda^{-1}a^{3^k}v)}\notag\\&=\sum_{a\in\mathbb{F}_{3^n}}\omega^{{\rm Tr}^{k}_{1}(\lambda^{-1}a^{3^k+1})\pm{\rm Tr}^{n}_{1}(\lambda^{-1}a^{3^k}v)+{\rm Tr}^{k}_{1}(\lambda^{-1}v^{3^k+1})}
\notag\\&=\sum_{a\in\mathbb{F}_{3^n}}\omega^{{\rm Tr}^{k}_{1}(\lambda^{-1}(a\pm v)^{3^k+1})}=-3^k
\end{align}
and
\begin{align}
&\sum_{a\in\mathbb{F}_{3^n}}\omega^{-{\rm Tr}^{k}_{1}(\lambda^{-1}a^{3^k+1})\pm{\rm Tr}^{n}_{1}(\lambda^{-1}a^{3^k}v)}=-3^k.
\end{align}
On the other hand, it follows from  ${\rm Tr}^{k}_{1}(\lambda^{-1}u^{3^k+1})=0$, ${\rm Tr}^{k}_{1}(\lambda^{-1}v^{3^k+1})=0$ and  ${\rm Tr}^{n}_{1}(\lambda^{-1}u^{3^k}v)=2$ that
\begin{align}
&\sum_{a\in\mathbb{F}_{3^n}}\omega^{{\rm Tr}^{k}_{1}(\lambda^{-1}a^{3^k+1})+{\rm Tr}^{n}_{1}(\lambda^{-1}a^{3^k}u)+{\rm Tr}^{n}_{1}(\lambda^{-1}a^{3^k}v)}
\notag\\=&\omega\sum_{a\in\mathbb{F}_{3^n}}\omega^{{\rm Tr}^{k}_{1}(\lambda^{-1}a^{3^k+1})+{\rm Tr}^{n}_{1}(\lambda^{-1}a^{3^k}u)+{\rm Tr}^{n}_{1}(\lambda^{-1}a^{3^k}v)+{\rm Tr}^{n}_{1}(\lambda^{-1}u^{3^k}v)}
\notag\\&\times\omega^{{\rm Tr}^{k}_{1}(\lambda^{-1}u^{3^k+1})+{\rm Tr}^{k}_{1}(\lambda^{-1}v^{3^k+1})}
\notag\\=&\omega\sum_{a\in\mathbb{F}_{3^n}}\omega^{{\rm Tr}^{k}_{1}(\lambda^{-1}(a+v+u)^{3^k+1})}
\notag\\=&-3^k\omega.
\end{align}
Similarly, we have
\begin{align}
\sum_{a\in\mathbb{F}_{3^n}}\omega^{{\rm Tr}^{k}_{1}(\lambda^{-1}a^{3^k+1})+{\rm Tr}^{n}_{1}(\lambda^{-1}a^{3^k}u)-{\rm Tr}^{n}_{1}(\lambda^{-1}a^{3^k}v)}
=&\omega^2\sum_{a\in\mathbb{F}_{3^n}}\omega^{{\rm Tr}^{k}_{1}(\lambda^{-1}(a-v+u)^{3^k+1})}
\notag\\=&-3^k\omega^2,
\end{align}
\begin{align}
\sum_{a\in\mathbb{F}_{3^n}}\omega^{{\rm Tr}^{k}_{1}(\lambda^{-1}a^{3^k+1})-{\rm Tr}^{n}_{1}(\lambda^{-1}a^{3^k}u)+{\rm Tr}^{n}_{1}(\lambda^{-1}a^{3^k}v)}
=&\omega^2\sum_{a\in\mathbb{F}_{3^n}}\omega^{{\rm Tr}^{k}_{1}(\lambda^{-1}(a+v-u)^{3^k+1})}
\notag\\=&-3^k\omega^2
\end{align}
and
\begin{align}
\sum_{a\in\mathbb{F}_{3^n}}\omega^{{\rm Tr}^{k}_{1}(\lambda^{-1}a^{3^k+1})-{\rm Tr}^{n}_{1}(\lambda^{-1}a^{3^k}u)-{\rm Tr}^{n}_{1}(\lambda^{-1}a^{3^k}v)}
=&\omega\sum_{a\in\mathbb{F}_{3^n}}\omega^{{\rm Tr}^{k}_{1}(\lambda^{-1}(a-v-u)^{3^k+1})}
\notag\\=&-3^k\omega.
\end{align}
Using  ${\rm Tr}^{k}_{1}(\lambda^{-1}u^{3^k+1})=0$, ${\rm Tr}^{k}_{1}(\lambda^{-1}v^{3^k+1})=0$ and  ${\rm Tr}^{n}_{1}(\lambda^{-1}u^{3^k}v)=2$ again, we obtain
\begin{align}
&\sum_{a\in\mathbb{F}_{3^n}}\omega^{-{\rm Tr}^{k}_{1}(\lambda^{-1}a^{3^k+1})+{\rm Tr}^{n}_{1}(\lambda^{-1}a^{3^k}u)+{\rm Tr}^{n}_{1}(\lambda^{-1}a^{3^k}v)}
\notag\\=&\omega^2\sum_{a\in\mathbb{F}_{3^n}}\omega^{-{\rm Tr}^{k}_{1}(\lambda^{-1}a^{3^k+1})+{\rm Tr}^{n}_{1}(\lambda^{-1}a^{3^k}u)+{\rm Tr}^{n}_{1}(\lambda^{-1}a^{3^k}v)-{\rm Tr}^{n}_{1}(\lambda^{-1}u^{3^k}v)}
\notag\\&\times\omega^{-{\rm Tr}^{k}_{1}(\lambda^{-1}u^{3^k+1})-{\rm Tr}^{k}_{1}(\lambda^{-1}v^{3^k+1})}
\notag\\=&\omega^2\sum_{a\in\mathbb{F}_{3^n}}\omega^{{\rm Tr}^{k}_{1}(-\lambda^{-1}(a-v-u)^{3^k+1})}
\notag\\=&-3^k\omega^2.
\end{align}
Similarly, we have
\begin{align}
\sum_{a\in\mathbb{F}_{3^n}}\omega^{-{\rm Tr}^{k}_{1}(\lambda^{-1}a^{3^k+1})+{\rm Tr}^{n}_{1}(\lambda^{-1}a^{3^k}u)-{\rm Tr}^{n}_{1}(\lambda^{-1}a^{3^k}v)}
=&\omega\sum_{a\in\mathbb{F}_{3^n}}\omega^{{\rm Tr}^{k}_{1}(-\lambda^{-1}(a+v-u)^{3^k+1})}
\notag\\=&-3^k\omega,
\end{align}
\begin{align}
\sum_{a\in\mathbb{F}_{3^n}}\omega^{-{\rm Tr}^{k}_{1}(\lambda^{-1}a^{3^k+1})-{\rm Tr}^{n}_{1}(\lambda^{-1}a^{3^k}u)+{\rm Tr}^{n}_{1}(\lambda^{-1}a^{3^k}v)}
=&\omega\sum_{a\in\mathbb{F}_{3^n}}\omega^{{\rm Tr}^{k}_{1}(-\lambda^{-1}(a-v+u)^{3^k+1})}
\notag\\=&-3^k\omega
\end{align}
and
\begin{align}\label{pp29}
\sum_{a\in\mathbb{F}_{3^n}}\omega^{{\rm Tr}^{k}_{1}(\lambda^{-1}a^{3^k+1})-{\rm Tr}^{n}_{1}(\lambda^{-1}a^{3^k}u)-{\rm Tr}^{n}_{1}(\lambda^{-1}a^{3^k}v)}
=&\omega^2\sum_{a\in\mathbb{F}_{3^n}}\omega^{{\rm Tr}^{k}_{1}(\lambda^{-1}(a-v-u)^{3^k+1})}
\notag\\=&-3^k\omega^2.
\end{align}
Combining (\ref{pp16})-(\ref{pp29}), we get
\begin{align*}
N_{0}=&3^{n-3}+\frac{1}{27}(-10\cdot3^k-4\cdot3^k(\omega+\omega^2))=3^{n-3}-\frac{1}{27}6\cdot3^k=3^{n-3}-2\cdot3^{k-2}.
\end{align*}
We next compute  $N_{1}$ and $N_{2}$.
According to   (\ref{pp13}),  if ${\rm Tr}^{k}_{1}(\lambda^{-1}a^{3^k+1})=2$, ${\rm Tr}^{n}_{1}(\lambda^{-1}a^{3^k}u)=0$ and ${\rm Tr}^{n}_{1}(\lambda^{-1}a^{3^k}v)=0$, then  $\widehat{\chi}_f(a)=-3^{k+1}\omega$. Therefore we have
\begin{align*}
N_{1}=&\frac{1}{27}\sum_{a\in\mathbb{F}_{3^n}}\sum_{x\in\mathbb{F}_{3}}\omega^{x({\rm Tr}^{k}_{1}(\lambda^{-1}a^{3^k+1})-2)}\sum_{y\in\mathbb{F}_{3}}\omega^{y{\rm Tr}^{n}_{1}(\lambda^{-1}a^{3^k}u)}\sum_{z\in\mathbb{F}_{3}}\omega^{z{\rm Tr}^{n}_{1}(\lambda^{-1}a^{3^k}v)}
\notag\\=&\frac{1}{27}\sum_{a\in\mathbb{F}_{3^n}}(1+\omega^{1+{\rm Tr}^{k}_{1}(\lambda^{-1}a^{3^k+1})}+\omega^{2-{\rm Tr}^{k}_{1}(\lambda^{-1}a^{3^k+1})})
(1+\omega^{{\rm Tr}^{n}_{1}(\lambda^{-1}a^{3^k}u)}
\notag\\&+\omega^{-{\rm Tr}^{n}_{1}(\lambda^{-1}a^{3^k}u)})
(1+\omega^{{\rm Tr}^{n}_{1}(\lambda^{-1}a^{3^k}v)}+\omega^{-{\rm Tr}^{n}_{1}(\lambda^{-1}a^{3^k}v)})
\notag\\=&\frac{1}{27}\sum_{a\in\mathbb{F}_{3^n}}
(1+\omega^{1+{\rm Tr}^{k}_{1}(\lambda^{-1}a^{3^k+1})}+\omega^{2-{\rm Tr}^{k}_{1}(\lambda^{-1}a^{3^k+1})})
\big(1+\omega^{{\rm Tr}^{n}_{1}(\lambda^{-1}a^{3^k}v)}
\notag\\&+\omega^{-{\rm Tr}^{n}_{1}(\lambda^{-1}a^{3^k}v)}
+\omega^{{\rm Tr}^{n}_{1}(\lambda^{-1}a^{3^k}u)}+\omega^{{\rm Tr}^{n}_{1}(\lambda^{-1}a^{3^k}u)+{\rm Tr}^{n}_{1}(\lambda^{-1}a^{3^k}v)}
\notag\\&+\omega^{{\rm Tr}^{n}_{1}(\lambda^{-1}a^{3^k}u)-{\rm Tr}^{n}_{1}(\lambda^{-1}a^{3^k}v)}
+\omega^{-{\rm Tr}^{n}_{1}(\lambda^{-1}a^{3^k}u)}+\omega^{-{\rm Tr}^{n}_{1}(\lambda^{-1}a^{3^k}u)+{\rm Tr}^{n}_{1}(\lambda^{-1}a^{3^k}v)}
\notag\\&+\omega^{-{\rm Tr}^{n}_{1}(\lambda^{-1}a^{3^k}u)-{\rm Tr}^{n}_{1}(\lambda^{-1}a^{3^k}v)}\big).
\end{align*}
By a similar process of calculating $N_{0}$, we can obtain
\begin{align*}
N_{1}=&3^{n-3}+\frac{1}{27}(-5\cdot3^k\omega-5\cdot3^k\omega^2-2\cdot3^k\omega(\omega+\omega^2)-2\cdot3^k\omega^2(\omega+\omega^2))
\\=&3^{n-3}+\frac{1}{27}(-3\cdot3^k(\omega+\omega^2))=3^{n-3}+3^{k-2}.
\end{align*}
According to Parseval equation $\sum_{a\in\mathbb{F}_{3^{n}}}|\widehat{\chi}_f(a)|^2=3^{2n}$, we get the number of  $a\in\mathbb{F}_{3^{n}}$ such that $|\widehat{\chi}_f(a)|= 3^{k+1}$ is equal to $3^{n-2}$ and the number of  $a\in\mathbb{F}_{3^{n}}$ such that $\widehat{\chi}_f(a)= 0$ is equal to $3^{n}-3^{n-2}$. This implies that $N_{2}=3^{n-3}+3^{k-2}$.


\end{document}